# The inverse proximity effect in strong ferromagnet-superconductor structures

V. O. Yagovtsev[1], N. G. Pugach[1,2] and M. Eschrig[3]

[1] National Research University Higher School of Economics, Moscow, 101000, Russia
[2] Skobeltsyn Institute of Nuclear Physics, Lomonosov Moscow State University, Leninskie Gory, 1 (2), Moscow GSP-2, 119991, Russia
[3] Institut für Physik, Universität Greifswald, D-17489 Greifswald, Germany

E-mail: vyagovtsev@hse.ru

## Abstract

The magnetization in a superconductor induced due to the inverse proximity effect is investigated in hybrid bilayers containing a superconductor and a ferromagnetic insulator or a strongly spin-polarized ferromagnetic metal. The study is performed within a quasiclassical Green function framework, wherein Usadel equations are solved with boundary conditions appropriate for strongly spin-polarized ferromagnetic materials. A comparison with recent experimental data is presented. The singlet to triplet conversion of the superconducting correlations as a result of the proximity effect with a ferromagnet is studied.

## 1. Introduction

In classical electronics, charge currents are used to transfer information, while in spintronics, spin-polarized currents are used for this purpose [1]. The heat, generated in the process of using spin-polarized currents, can be an undesirable spurious effect. The use of superconductors in magnetic nanostructures can greatly reduce this heating, increasing the energy efficiency of spintronic devices. This idea underlies superconducting spintronics.

In the last 30 years, superconducting spintronics has been actively studied by many experimental and theoretical groups [2–6]. The main focus of the research was the theoretical and experimental description of hybrid nanostructures. In particular, the proximity effect in superconductor-ferromagnet (SF) structures was investigated. Its main features are spatial oscillations of the amplitude of the superconducting correlation penetrating into the ferromagnet [4,5,7–9] and the appearance of triplet superconducting correlations with inhomogeneous magnetization of the layers [6,10]. It leads to a long-range proximity effect [5,6]. This was discovered in experimental studies and it is now known that the long-range proximity effect leads to change in the critical temperature of superconductors in bilayer and multilayer SF structures, oscillations of the Josephson current in the presence of a ferromagnet, and the long-range Josephson effect [2–5]. Devices of superconducting spintronics are considered as promising to create sensitive sensors and element base for a quantum computer [5,11].

The inverse proximity effect in a superconductor in contact with a ferromagnet was first described in the work [12]. Triplet superconducting correlations are the origin of the non-zero induced magnetization in the superconductor due to inverse proximity effect. Triplet Cooper pairs are symmetric in spin space [13–16]. In theoretical work [17] the induced magnetization in such structures was estimated in the presence of long-range triplet superconducting correlations. In experimental studies [18,19] evidence of the presence of induced magnetization in the proximity of a superconductor to a ferromagnetic metal was demonstrated.

The origin of the magnetization is a spin-splitting of the local density of states in the superconductor near the SF interface. This spin-splitting is directly related to the spin-polarization of the Cooper pair. As a result, they will create a noticeable magnetic field, which has the feature that it depends on the concentration of Cooper pairs. In the absence of superconductivity, this magnetic field will be restricted to a much shorter length scale, related to the Thomas-Fermi length [20].

Another way to explain the appearance of magnetization is the spin mixing angle [21]. It quantifies how large the relative scattering phase shift between the electrons of the Cooper pair is after they are reflected from an S–F interface. This difference between phase of the spin-up and spin-down electrons of the Cooper pair results in an imbalance in the DOS for electrons with different spin, and it leads to the appearance of an induced magnetization.

However, most of these theoretical and experimental studies were focused on structures with ferromagnetic metals (FM). Later phenomena associated with the inverse proximity effect in structures with ferromagnetic insulators (FI) were experimentally demonstrated [21,22]. In addition, in experiments [23–25] phenomena related to the proximity effect of such structures were demonstrated. They include splitting in the density of states of the superconductor due to the effective exchange field created by the proximity with a ferromagnetic insulator [23] and the possibility of manipulating spin transfer by adding ferromagnetic insulators to the superconducting layer [25].

The present work is devoted to the theoretical description of induced magnetization in these structures. In Section 3, a comparison is made with experimental studies [25–27], where the value of the induced magnetization is obtained. In the theoretical model of induced magnetization in S-FI and S-FI-S structures [28,29] the Hamiltonian consists of a general Hamiltonians of the S layers and a tunnelling term with spin-dependent matrix elements. The latter represent the band splitting in the ferromagnetic barrier which leads to different tunnel barrier heights for spin-up and spin-down electrons, and consequently, to differences between tunnelling amplitudes for electrons with different spin projections. Electrons with spin-up and spin-down will penetrate the insulating barrier of the S-FI bilayer to different depths. In both cases this will lead to a spin imbalance in the S layer near the S-FI interface and produce spin magnetization in the superconductor. And again, absence of Cooper pairs will result in absence of this mechanism for induced magnetization.

Our work also describes induced magnetization in structures with strongly ferromagnetic metals (Section 2.3). The work is based on the quasiclassical approximation using Usadel equations with boundary conditions appropriate for the case of strong spin polarization of the ferromagnet [21,30]. These equations are applicable in the dirty limit of a superconductor, in which the mean free path of an electron is much less than the coherence length of a Cooper pair.

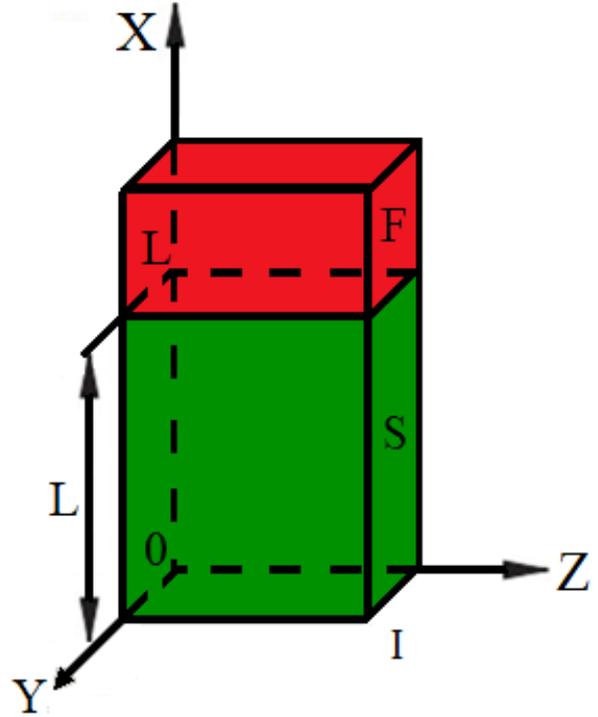

**Figure 1.** The schematic view of the simulated superconducting-ferromagnet structure.

The structure we describe is shown schematically in Fig. 1. Layer S is the superconductor layer, and F is a ferromagnetic insulator or ferromagnetic metal. Here $L$ is the thickness of the superconducting layer.

Section 2.1 provides a model describing the S-F bilayer using linearized Usadel equations. The solution to these equations serves as the basis for calculating the magnetization induced in the superconductor. Section 2.2 shows boundary conditions and results for the induced magnetization for the S-FI case, Section 2.3.1 shows them for the S-FM case at the temperature close to its critical value and Section 2.3.2 presents the equations for S-FM case in the limit of weak proximity effect.

Section 3.1 presents the determination of the angle of spin mixing from comparison with the theoretical model [31], and Section 3.2 presents the determination of the angle using related experimental data [25–27]. Section 3.3 is devoted to the results of the numerical calculations and shows how the magnetization depends on various parameters of the model. Section 4 discusses the work's main results.

## 2. Model

### 2.1. Equations for superconductor in contact with ferromagnet

To estimate the value of the induced magnetization in the superconductor we use the quasiclassical approximation in the form of Usadel equations, assuming that the mean free path is the smallest characteristic length in all layers of our structure. In the ferromagnet, uniform magnetization is assumed over the entire volume of the material. It is assumed that the diffusive limit is realized in the structure, as real superconducting nanostructures created by sputtering usually satisfy the conditions of the diffusive, or "dirty" limit. The coherence length in the superconductor is $\xi = \sqrt{\hbar D_s/2\pi k_B T_{cb}}$, where $\hbar$ is the reduced Planck constant, $k_B$ is the Boltzmann constant (we use units where $\hbar = k_B = 1$), $D_s$ is the electron diffusion coefficient of the superconductor, and $T_{cb}$ is the critical temperature of the bulk superconductor.

The linearized Usadel transport equations within the superconductor, applicable in the dirty limit [32] with temperature in the vicinity of the S film's critical superconducting temperature $T_c$, are:

$$(D_s\nabla^2 - 2|\omega_n|)f_s = -2\pi\Delta,$$
$$(D_s\nabla^2 - 2|\omega_n|)\mathbf{f}_t = 0. \quad (1)$$

Here the Matsubara frequencies are $\omega_n = \pi T(2n+1)$, $n$ is an integer, T is the temperature, $f_s$ and $\mathbf{f}_t$ are the singlet and triplet components of the anomalous Green function, respectively, and $\mathbf{f}_t$ is the vector $(f_{tx}, f_{ty}, f_{tz})$, $\Delta$ is the scalar superconducting order parameter which we assume to be real-valued, and in the bulk equal to the superconducting energy gap. As a result of the influence of the ferromagnetic layer, the singlet superconducting correlations near the interface with the insulator is suppressed and converted to the triplet ones, its value depends on the distance to the S-F interface. The F layer magnetization is considered to be aligned along the z axis.

The boundary condition at the boundary of the superconductor with the (outer) environment has the form

$$\frac{\partial f_s}{\partial x} = \frac{\partial f_{tz}}{\partial x} = 0,$$

reflecting zero current flow through this interface.

In both cases of FI and FM, after finding the Green's functions by solving the Usadel equation with boundary conditions, it is necessary to take into account the suppression of the superconducting correlations. The suppression of the singlet component at the interface in the limit of small transparency of the S-F interface goes like the square of the spin-mixing angle, and can thus be neglected because we are interested only in effects of linear order in the spin-mixing angle [33]. To show this, we calculate the singlet and triplet pair amplitudes and make sure that the order parameter has almost constant value. In order to do this, the self-consistent equation [15]

$$\Delta(x)\ln\frac{T_c}{T} = \pi T \sum_{n>0}\left(\frac{\Delta(x)}{\omega_n} - \frac{f_s(x)}{\pi}\right). \quad (2)$$

Is also solved approximately in the appropriate case. The induced magnetization is calculated by the formula [15]:

$$\delta M(x) = 2\mu_B N_0 \pi k_B T \sum_{n>0} \mathbf{g}(x, \omega_n), \quad (3)$$

where $\mu_B$ is the Bohr magneton, $N_0$ is the density of states at the Fermi level in the normal state.

Equations (1) are supplemented by the boundary condition for the boundaries parallel to the YOZ plane. In our work, we consider boundary conditions appropriate for a ferromagnetic insulator and for a ferromagnetic metal.

### 2.2. Bilayer with ferromagnetic insulator

In this case the boundary condition [21,30] at the boundary of the S and FI layers is:

$$A\sigma_S g_0 \frac{\partial (f_{tz} + f_s)}{\partial x} = -2NG_Q P(\varphi), \quad (4)$$

where $A$ is the contact area, $\sigma_S$ is the conductivity of the superconductor in the non-superconducting state, $g_0$ is the singlet component of the Green function g, $N$ is the number of conduction channels at the boundary (in the general case, it is determined by the polarization of the ferromagnet, the coefficient of particle transmission through the boundary, and the conductivity of the boundary), $G_Q = e^2/\pi$ is the quantum of conductivity. $P(\varphi)$ is the function defined in [21], which depends on the value of $\varphi$, the angle of spin mixing.

The spin mixing angle describes how strongly the magnetic exchange field affects the phase difference for electrons with spin-up and spin-down at the superconductor-ferromagnet interface. Since the ferromagnetic insulator may be considered [28,29] as a spin-dependent potential barrier, spin-up and spin-down electrons penetrate into FI at different depths. The simplest approach leads to the assumption that the stronger the magnetic exchange field in the ferromagnetic material, the greater the difference in penetration depth, and therefore, the phase difference created by the boundary will be larger. However, it was shown [34] that the dependence of the created phase difference on the exchange splitting is nonlinear at large splitting. The angle of spin mixing $\varphi$ can be estimated using the model given in the work [31].

The vector component of the normal Green function **g** is expressed through the combination of the components of the anomalous Green function,

$$f\tilde{f} + g^2 = \sigma_0.$$

Here $\sigma_0$ is a unit matrix, $g = g_0 + (\mathbf{g} \cdot \boldsymbol{\sigma})$, $f = [f_s + (\mathbf{f}_t \cdot \boldsymbol{\sigma})]i\sigma_y$, where **g** is the spin vector part of the normal Green function g, $\boldsymbol{\sigma}$ is the vector of spin Pauli matrices.

A linear approximation was used, which is justified by a small change in the normal (diagonal part of the Nambu-

Gor'kov matrix) Green function and the smallness of the anomalous (off-diagonal) Green function. To obtain an analytic solution of the Usadel equation with two boundary conditions, the limit of temperature close to the critical temperature was taken.

After expanding the normal Green function g into a Taylor series over the anomalous Green function components and discarding the terms of third and higher order using equation (3), one obtains the following expression for the induced magnetization [17]:

$$\delta M(x) = -4\mu_B N_0 \pi k_B T \sum_{n>0} \text{Im}\left(f_s(x,\omega_n)\tilde{f}_{tz}(x,\omega_n)\right). \quad (5)$$

Note that neither a pure singlet nor a pure triplet pair can create such a spin-splitting, only the simultaneous presence of both, see discussion around Eq. (87) in review [5].

We assume that it is possible to neglect the Meissner effect for this structure, which acts only at lengths of the order of the London penetration depth $\lambda$, [35] i.e. it is assumed that the thickness of the S layer is $L \sim \xi_S \ll \lambda$.

## 2.3. Ferromagnetic metal case

### 2.3.1 Limit of temperature close to the critical temperature

Now we turn to the case of a ferromagnetic metal (FM) as magnetic element in the bilayer. We consider the case of a strongly spin-polarized ferromagnet where the period of space oscillations of the superconducting correlations and its decay length are much smaller than the mean free path $l_F$ and the layer thickness $L$. It means that the exchange magnetic energy $H$ is large enough to satisfy the inequality $\xi_H \ll l_F, L$, where $\xi_H = \hbar v_F/H$ for the magnetic length, where $v_F$ is Fermi velocity in the ferromagnet. The anomalous Green function describes the superconducting correlations, which penetrate in the F layer due to the proximity effect. The mean value of the short-range components of the anomalous Green function is negligible due to its fast oscillations in the strong ferromagnet.

Thus, on a length-scale considerably larger than the mean free path, we may assume that the short-range components of the anomalous Green function are zero and normal Green function is a constant in the ferromagnet. In this case we can consider the ferromagnet as semi-infinite and focus our attention on the S layer with appropriate boundary conditions at the SF interface. We assume that the ferromagnet occupies the semi-space $x > L$. For the same reason as for S-FI case, we assume that $L$ is much less than the London penetration depth $\lambda$. We also consider limit of temperature close to the critical temperature.

The Usadel equation in the superconductor has the similar form as (1),

$$\left(\frac{D_s}{\pi}g\frac{d^2}{dx^2} + 2i\omega\right)(f_s \pm f_{tz}) = -2\Delta g,$$

but the boundary condition at the boundary of the S and FM layers is different from the S-FI case. For the ferromagnetic metal the boundary condition [36] at the interface ($x = L$) may be written in the simple closed form

$$\frac{\hbar A \sigma_S}{e^2} \frac{\pi \omega_n}{\sqrt{\omega_n^2 + \Delta^2}} \frac{\partial (f_s \pm f_{tz})}{\partial x} = (t \mp i\varphi)(f_s \pm f_{tz}), \quad (6)$$

where $t = t_\uparrow + t_\downarrow$ is the probability for the particles to be transmitted through the S-FM interface. Here, $t_\uparrow = \sum_{n_\uparrow}|t_{\uparrow\uparrow n}|^2$, $t_\downarrow = \sum_{n_\downarrow}|t_{\downarrow\downarrow n}|^2$ are the transmission probability matrix elements for each of the $n_\uparrow$ or $n_\downarrow$ transmission channels with spin up (↑) or down (↓), respectively [36]. We assume no spin-flip scattering on the SF interface, therefore the transmission matrix has a diagonal form. There are different numbers of transmission channels for spin-up and spin-down in the ferromagnet, therefore $n_\uparrow \neq n_\downarrow$.

The imaginary part $i\varphi$ appears due to the spin splitting in the ferromagnet. For a structure without Josephson current, when $\Delta$ may be assumed to be real, $(f_s - f_{tz}) = (f_s + f_{tz})^*$. Earlier calculations of the proximity effect between a superconductor and a strong ferromagnet [37,38] were performed for only a very thin F layer $L \sim \xi_H$. In this case the boundary conditions at the SF interface had approximately the same form [38] with a complex coefficient between the functions $(f_s \pm f_{tz})$ and their derivatives, but these coefficients oscillated with thickness $d_F$.

The inverse proximity effect leads to the singlet to triplet transform of the superconducting correlations near the SF interface. If $L \sim \xi$, the order parameter and the anomalous Green function $f_s + f_{tz} = \frac{\pi \Delta(x)}{|\omega_n| + D_s k^2/2}$ may be determined by the ansatz [38]

$$\Delta(x) = \Delta_0 \cos[k(x - L)]. \quad (7)$$

This ansatz satisfies the Usadel equation (1) with the boundary condition at the free interface. For this case, for the dependence of all the quantities of interest, we will concentrate only on thicknesses of the order of the coherence length, since the ansatz does not work well at larger thicknesses.

The value of $k$ is found from the boundary condition (6) that leads to the equation

$$k \cdot \tan(kL) = \frac{e^2}{2\pi^2 \hbar A \sigma_S}(t - i\varphi). \quad (8)$$

Note that here k depends on the values $L$, $t$ and $\varphi$, and does not depend on $\omega_n$. Using the obtained value of $k$ we may calculate properties of the proximity structure (consisting of a superconductor – strongly spin-polarized ferromagnet bilayer) as functions of the parameters of the SF boundary, $t$ and $\varphi$.

As it was mentioned above, the inverse proximity effect between a ferromagnet and a superconductor leads to an induced magnetization at the S side [14]. Such induced magnetization was detected in experiments using the nuclear magnetic resonance [19], and the polar Kerr effect [39]. The vector of this induced magnetization may be found by the formula (3). Using this formula, we calculated the magnetization at the S-FM interface depending on the thickness of the superconductor $L$.

## 2.3.2 The weak proximity effect limit

Here we assume that the transparency of the SF interface is small, and therefore the Green functions weakly differ from their bulk values $g = \frac{-i\omega_n}{\sqrt{\omega_n^2 + \Delta^2}}$, $f = \frac{\Delta}{\sqrt{\omega_n^2 + \Delta^2}}$ and $f_{tz}, f_s$ are defined by the boundary condition (6).

The corrections to the bulk values of functions $g$ and $f$ appear only in the second order on $f_{tz}$ and may be neglected. The solution of the Usadel equation (1) satisfying boundary condition at the free interface has the form $f_{tz} = C \cosh k_S(x - L)$, where $k_S = 2\sqrt{\omega_n^2 + \Delta^2}/\hbar D_s$. The constant $C$ may be found from the boundary condition at the SF interface (6),

$$C = if\varphi \, sign(\omega_n) \left( \frac{\pi A \sigma_S}{e^2} \frac{2|\omega_n|}{D_s k_S} \sinh k_S L - t \cosh k_S L \right)^{-1}.$$

As it is expected, the amplitude $C$ of the triplet component is proportional to the phase shift $\varphi$, which the anomalous function experiences at the interface with the ferromagnet. This value depends on the exchange field $H$ and is equal to zero in the absence of $H$. The expression for the induced magnetization (5) may be written as

$$\delta M(x) = -4\mu_B N_0 \pi k_B T \cdot$$
$$\cdot \sum_{\omega_n > 0} \frac{\Delta^2}{\omega_n \sqrt{\omega_n^2 + \Delta^2}} \frac{\varphi \cosh[k_S(x-L)]}{\frac{\pi A \sigma_S}{e^2} \frac{2\omega_n}{D_s k_S} \sinh k_S L - t \cosh k_S L}.$$

For a thick superconductor for $L \gg \xi$ we may write this equation in the form:

$$\delta M(x) = -4\mu_B N_0 \pi k_B T \sum_{\omega_n > 0} \frac{\Delta^2}{\omega_n \sqrt{\omega_n^2 + \Delta^2}} \frac{\varphi \exp(-k_S x)}{\frac{\pi A \sigma_S}{e^2} \frac{2\omega_n}{D_s k_S} - t}.$$

The induced magnetization decreases exponentially from the SF boundary at a distance of the order of the superconducting coherence length $\xi$ at low temperature.

## 3. Results

### 3.1. Comparison with experiment

The spin mixing angle can be estimated by comparing the magnetization obtained for different values of the angle with the magnetization which was taken from experimental studies [25–27]. In these works, the conductivity of the structures under study was measured, and the densities of states (DOS) were obtained. The values of the exchange field in the superconductor were obtained from DOS.

Table 1 shows values of the spin-mixing angle that were obtained from experimental studies as a result of fitting the data. In these works, Al was taken as a superconductor and EuS was taken as a ferromagnetic insulator. $\sigma_S$ was taken as $3.8 \cdot 10^7 \, Ohm^{-1} \cdot m^{-1}$. If it were not mentioned in the article, $D_s$ was taken as $8.68 \cdot 10^{-3} s^{-1} \cdot m^2$, because we can assume that its value in all articles is approximately the same due to almost identical production conditions.

| № | L, nm | $D_s$, $s^{-1} \cdot m^2$ | B, T | φ |
|---|---|---|---|---|
| 22 | 10 | $2 \cdot 10^{-3}$ | 1.75 | 0.033 |
| 23 | 10-15 | $8.68 \cdot 10^{-3}$ | 2.25-4.5 | 0.14-0.27 |
| 24 | 20 | $8.68 \cdot 10^{-3}$ | 0.25 | 0.8 |

**Table 1.** Comparison with the experimental works.

One can see that the spin mixing angle does not reach its maximum value π/2. This is consistent with the theoretical dependence of the magnetization on the spin mixing angle.

The model calculation fit with the experimental data yields some variation of the spin mixing angle for different superconductor thicknesses, although the spin-mixing angle depends only on the properties of the interface between materials. Probably this is a consequence of the constant order parameter approximation, which nevertheless allows estimation of φ.

### 3.2. Estimation of the angle of spin mixing

To determine the spin mixing angle, we used a comparison with the boundary condition [31] written in a linearized form. After carrying out the corresponding transformations, we obtain the following relationship between the phenomenological parameters:

$$\sin(\varphi) = \frac{-A g_0}{2 N G_Q R_B L} \frac{G_\Phi}{G_T},$$

Here, $R_B$ is the barrier resistance, which is a phenomenological parameter, and $G_\Phi$ and $G_T$ are the phenomenological parameters given in the work [31].

The following estimates were obtained for the spin mixing angle in radians: $\varphi = 0.82$ at $N/A = 2.48 \cdot 10^{19}$ μm$^{-2}$, $T = 0.8 \cdot T_c$, $T_c = 1.2$ K, $L = 5\xi$, $\xi = 93$ nm.

Parameters appropriate for aluminium are taken for the superconductor. Aluminium has $T_c = 1.2$ K, and in the currently used preparation technology $\sigma = 3.8 \cdot 10^7$ Ohm$^{-1} \cdot$ m$^{-1}$, $D = 8.68 \cdot 10^{-3}$ s$^{-1} \cdot$ m$^2$. Parameters appropriate for cobalt are taken for the ferromagnet metal in our work.

### 3.3. Numerical results and discussion

The magnetization profile was obtained over the thickness of the superconducting layer as a function of the spin mixing angle. The structure was modelled with the following parameters: $N/A = 2.48 \cdot 10^{19}$ μm$^{-2}$, $T = 0.8 \cdot T_c$, $T_c$ is 1.2 K, $\sigma = 3.8 \cdot 10^7$ Ohm$^{-1} \cdot$ m$^{-1}$, $D = 8.68 \cdot 10^{-3}$ s$^{-1} \cdot$ m$^2$, $L = 4\xi, \xi = 93$ nm, $t = 0.7$. All dependencies are presented for given parameter values, unless otherwise indicated.

The calculation of pairing amplitudes was carried out: the values of Green functions with a constant order parameter were substituted into the equation (2). The result is presented in Fig. 2. It allows estimation of the suppression of the superconducting correlations at the boundary. It is seen that the superconducting correlations, and consequently, the order parameter of the superconductor is weakly suppressed by the magnetization of the neighbouring insulator layer at small spin mixing angles.

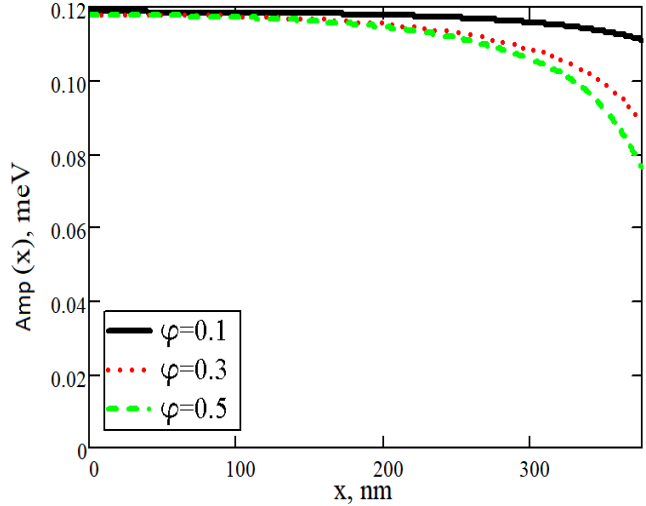

**Figure 2.** Pairing amplitudes of the superconductor for S-FI case as a function of the coordinate for various values of the angle of spin mixing, $\varphi$.

The obtained dependences of the components of the anomalous Green function are presented in Fig. 3.

The relations between the various components of the anomalous Green function are:

$f_S = \frac{f_{\uparrow\downarrow} - f_{\downarrow\uparrow}}{2}, f_{tz} = \frac{f_{\uparrow\downarrow} + f_{\downarrow\uparrow}}{2}$.

The dependences of the components $f_S$ and $f_{tz}$ of the anomalous Green function on the x coordinate, shown in Fig. 3, demonstrate that the singlet component of the Green function is suppressed when approaching the F boundary. The triplet component, on the contrary, decreases when moving away from the S-FI boundary. This corresponds to physical expectations, according to which, singlet Cooper pairs are transformed into triplet pairs at the S-FI border.

The functions shown in fig. 3 were substituted into formula (5). As a result, the induced magnetization values were obtained.

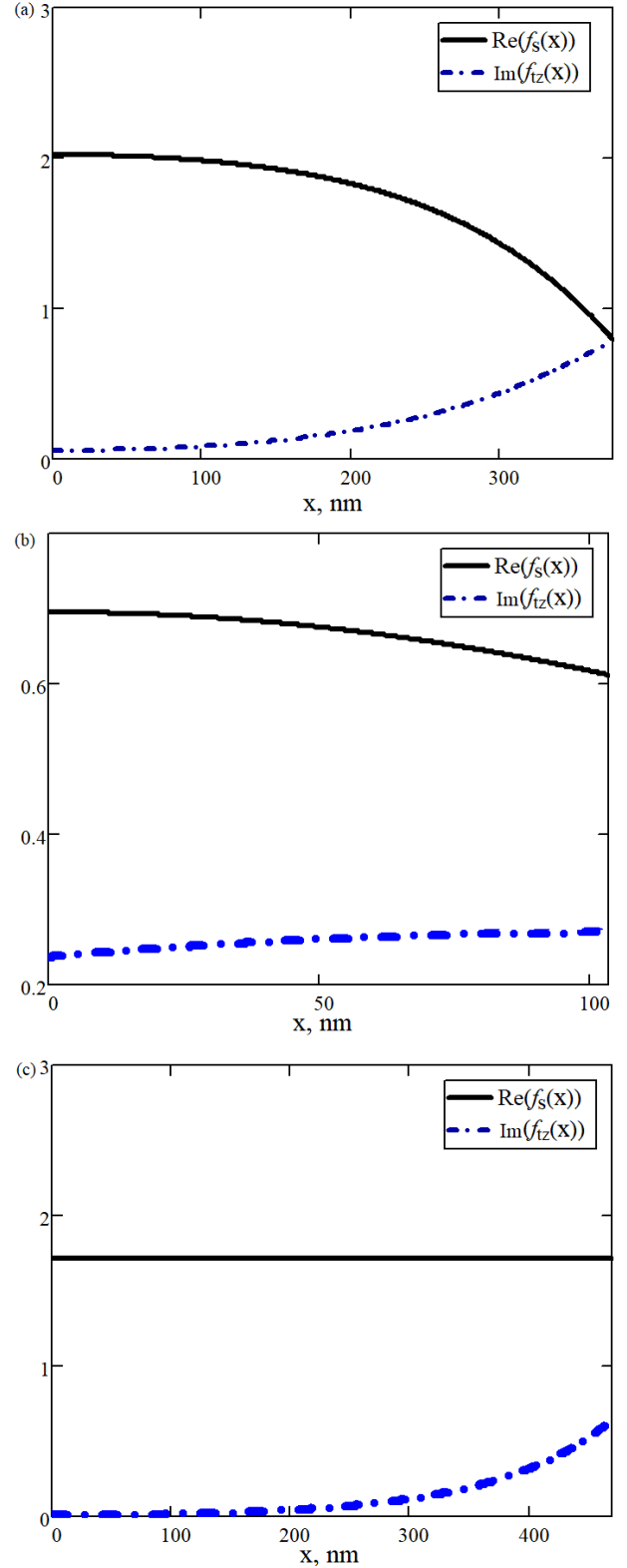

**Figure 3.** Real and imaginary components $f_s$ and $f_{tz}$ of the anomalous Green functions in the superconductor as functions of the coordinates, of (a) S-FI interface, (b) S-FM interface for the temperature close to the critical one, (c) S-FM interface for the weak proximity effect.

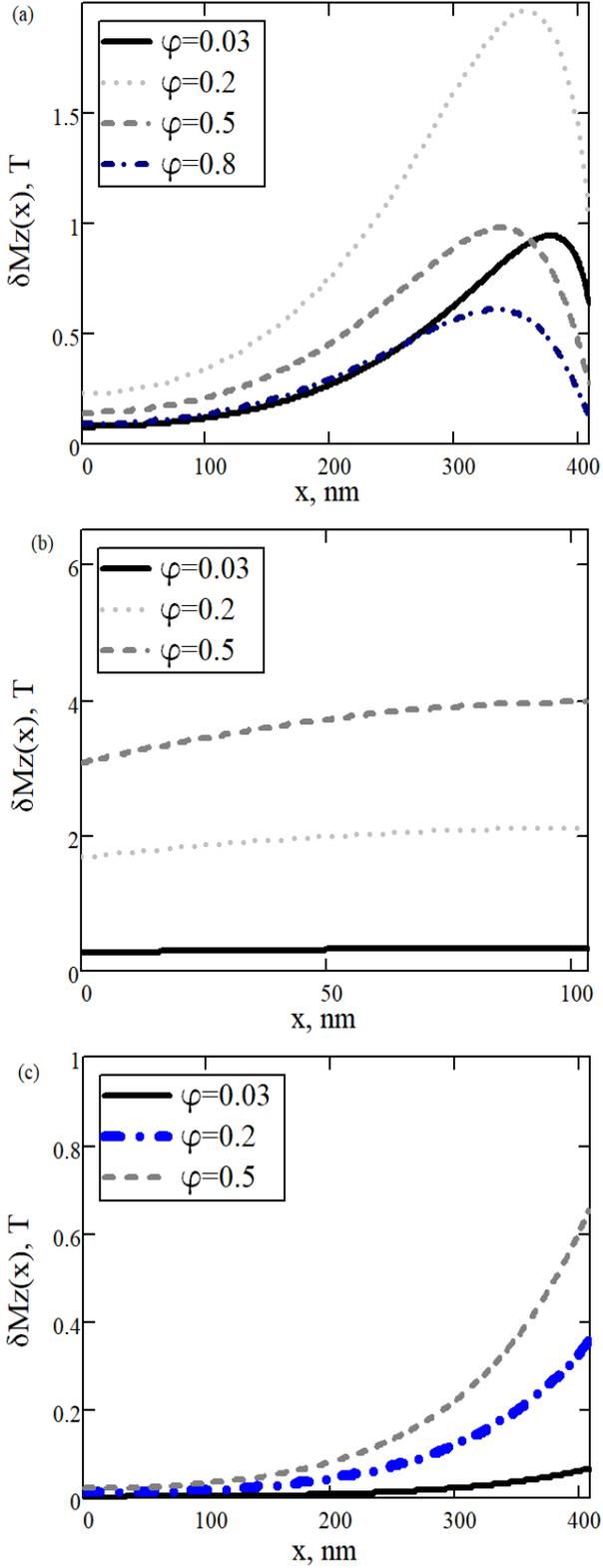
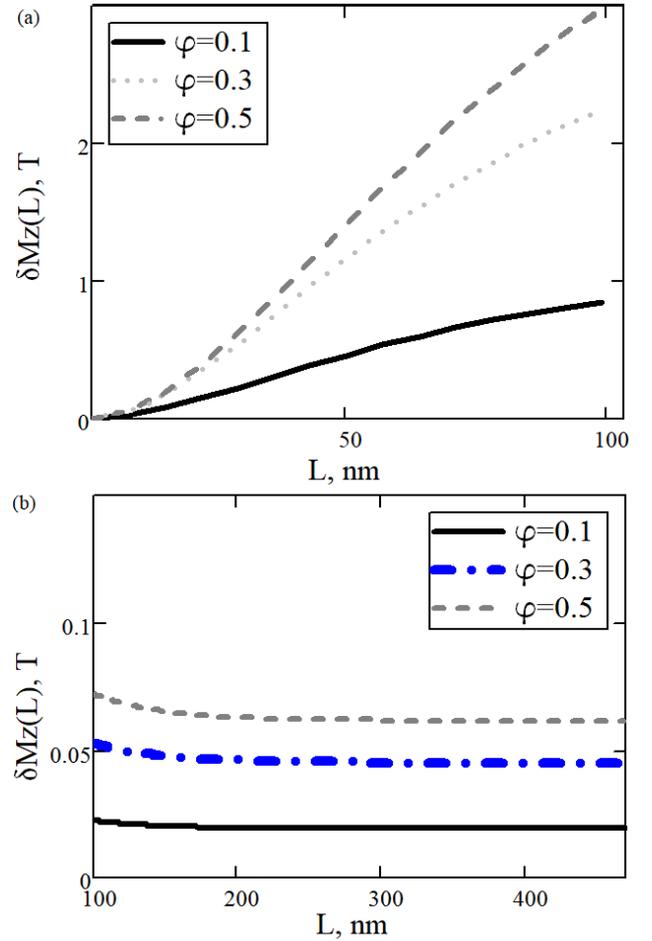

As it can be seen from Fig. 4(a), a local maximum of magnetization appears on the curve at the extremal value of the coordinate x. It becomes clear that such a behaviour of magnetization should have been expected. The reason for this behaviour is the decrease of the singlet component and the growth of the triplet component when approaching the superconductor-ferromagnetic interface.

We can see from Fig. 4 (b, c) the absence of a local extreme point of the induced magnetization profile at a contact with a ferromagnetic metal. We may assume that this is a consequence of the difference in boundary conditions, describing a nonzero transparency of the interface for electrons in the metallic case. The S-FM interface provides not only an inverse, but also a direct proximity effect with drainage of Cooper pairs from S into the ferromagnet. The S-FI interface produces only singlet-to-triplet conversion of Cooper pairs. This is the reason for different behaviour of the induced magnetization in the superconductor in the vicinity of these interfaces.

**Figure 4.** Magnetization versus the *x* coordinate for various values of the angle of spin mixing, $\varphi$ in the S layer of (a) S-FI bilayer, (b) S-FM bilayer for the temperature close to the critical limit, (c) S-FM bilayer for the weak proximity effect limit.

**Figure 5.** Magnetization in the S layer near the S-FM interface versus the superconductor thickness *L* for various values of the angle, $\varphi$, (a) for the temperature close to the critical limit, (b) for the weak proximity effect limit, here $T = 0.8 \cdot T_c$, $t = 0.7$ for (a), $t = 0.1$ for (b).

Fig. 5 (a) shows that, as one would expect, the magnetization at the S-FM boundary for small L is minuscule, because superconductivity is suppressed by the magnetization of a ferromagnet. The higher the angle, the less magnetization occurs. It is clear that growth of the ferromagnetic insulator's magnetization increases the suppression of superconductivity in the S-layer.

For the S-FM case within weak proximity effect, see Fig.5 (b). In this case, for small thickness of superconductor the magnetization is bigger than for much larger thickness. However, this might be due to the fact that in this case the suppression of the order parameter in the superconductor layer was not taken into account, unlike in case (a).

From Fig. 6(a), we see that there is a maximum of the average magnetization on the spin mixing angle $\varphi$ due to the following factors. For a small magnetization of the ferromagnet, the growth of the spin-mixing angle leads to the appearance of more triplet pairs, and the magnetization in a superconductor increases. However, at some point, the magnetization of the ferromagnet reaches such a value that its growth already leads to a decrease in the number of Cooper pairs due to the suppression of superconductivity, and, as a consequence, to a decrease in the number of singlet pairs from which triplet pairs are obtained. In addition, it is seen that with increase in temperature, the magnetization decreases. This is due to the fact that the superconductivity is suppressed with increasing temperature and, as a result, fewer singlet and triplet Cooper pairs are generated and contribute to the magnetization. There is no maximum of the average induced magnetization function $IM(\varphi)$ at Fig.6 (b,c) for the S-FM bilayer. This is also a consequence of the transparency of the interface.

## 4. Conclusion

We have presented results for the induced magnetization and superconducting correlations profile in the superconducting layer of a bilayer superconductor-ferromagnet structure. We have shown that the induced magnetization monotonically decreases with increasing temperature as a consequence of the inverse proximity effect which vanishes at the critical temperature. The dependence of the induced magnetization on the coordinate and on the spin mixing angle, has a maximum for S-FI, and is monotonous for S-FM structures. The reason for this behaviour is a decrease in the singlet component and a growth of the triplet component when approaching the superconductor-ferromagnetic interface. As the magnetization is proportional to the product of both, a local maximum in the profile appears.

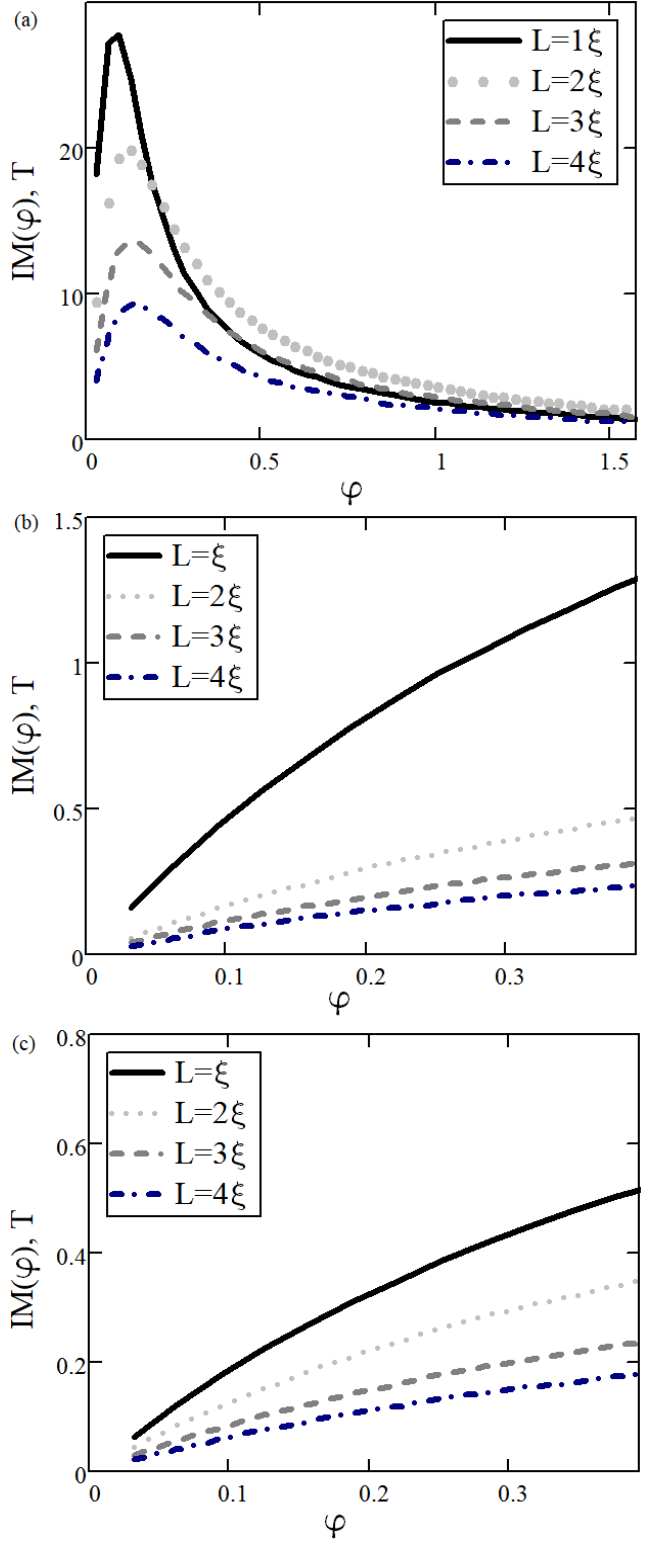

**Figure 6.** Average magnetization at (a) the S-FI bilayer, (b) the S-FM interface for the temperature close to the critical limit, (c) the S-FM interface for the weak proximity effect limit, versus the angle of spin mixing, $\varphi$, for various values of the temperature, $T$. Here $\xi = 93$ nm. The transparency $t = 0.7$ (b), $t = 0.1$ (c).

The reason for the local maximum in the averaged magnetization as function of the spin mixing angle is the competition between singlet-to triplet conversion, provided by spin-mixing. A stronger spin mixing leads to suppression of superconductivity and, therefore, to a decrease of the induced magnetization.

The magnitude of the induced magnetization is also different for S-FI and S-FM structures. In the S-FI case it is usually larger as a consequence of the finite transparency of the S-FM interface for electrons.

Superconductor-ferromagnet structures as discussed in this paper may have various applications, e.g. as sensitive bolometers and thermometers [40,41].

## Acknowledgements

Pugach N. and Yagovtsev V. thank Prof. Ekomasov E. G. for valuable discussions and the program Mirror Labs of National Research University Higher School of Economics for support of interrussian cooperation.

The calculations of inverse proximity effect in S-FI bilayers and estimations of the spin-mixing angle were funded by the Russian Ministry of Education and Science, Megagrant project N 2019-220-07-6383. The study of S-FM structures and comparison with experiment on inverse proximity effect were funded by RFBR according to the research project № 19-02-00316.